\documentstyle[preprint,revtex]{aps}

\begin{document}
\thispagestyle{empty}


\begin{center}
\hfill{IP-ASTP-02-95}\\
\hfill{May 1995}\\
\hfill{(Revised)}

\vspace{1 cm}

\begin{title}
A Calculation of Cosmic Variance in CMB Anisotropy\\

\end{title}
\vspace{1 cm}

\author{Kin-Wang Ng and A.~D.~Speliotopoulos}
\vspace{0.5 cm}

\begin{instit}
Institute of Physics, Academia Sinica\\ Taipei, Taiwan 11529, R.O.C.
\end{instit}
\end{center}
\vspace{0.5 cm}

\begin{abstract}
\vspace{1 cm}

\noindent We present a theoretical calculation of the variance
$\Delta C_l$ of the CMB anisotropy power spectrum $\langle
C_l\rangle$ caused by gravitational waves based on
quantum field theory in an inflationary cosmology.

\vspace{1 cm}
\end{abstract}
\newpage

One of the most exciting discoveries in astrophysics in the past few
years has been the presence of an anisotropy in the cosmic microwave
background (CMB) $\cite{COBE}$. It is generally believed that this
anisotropy is induced by metric fluctuations generated in the early universe.
CMB anisotropy measurements could thus provide an invaluable test of
their origin. However, before one can really compare the theoretical
prediction of the induced anisotropy in any specific model to the
experimental data, one has to know within the model the theoretical
error of the calculation. This error, which is often called cosmic
variance, is associated with the underlying statistics of the
fluctuations. While a theoretical calculation gives a definite
ensemble-averaged mean
value for the anisotropy, a terrestrial measurement only measures
the anisotropy of a single sample universe. If one uses the
experimental data to constrain a model, one will then have to take
into account the theoretical error of this mean value.

In CMB measurements, the measured temperature anisotropy $\delta T/T
(\hat e)$ (where $\hat e$ is a unit vector pointing to the celestial
sphere) is usually expanded in terms of spherical harmonics, $\delta
T/T = \sum_{l,m} a_{lm} Y_{lm}$, from which a two-point temperature
correlation function can be constructed. After
averaging over the celestial sphere (sky-averaging), the function is given by
\begin{equation}
C(\theta) \equiv \langle\frac{\delta T}{T}(\hat e_1)\frac{\delta T}{T}(\hat
e_2)\rangle_{sky} = \frac{1}{4\pi}\sum_l C_l P_l(\cos\theta)\>,
\label{ct}
\end{equation}
where $\theta$ is the separation angle, and $C_l\equiv \sum_{m=-l}^l
a^\dagger_{lm} a_{lm}$ is the anisotropy power spectrum.

Estimates of this cosmic variance in the anisotropy power spectrum
have been made previously
$\cite{Wise1,Wise2,Bond,Krauss,White,Turner}$. They claim that all
$a_{lm}$'s are independent gaussian random  variables satisfying
\begin{equation}
\langle a^\dagger_{lm} a_{l'm'}\rangle_{ens} =
\frac{\langle C_l\rangle_{ens}}{2l+1}\delta_{ll'}\delta_{mm'}\>,
\label{o}
\end{equation}
where $\langle\rangle_{ens}$ denotes taking ensemble average. As there are
$2l+1$ of the $a_{lm}$'s for each $l$, using chi-squared statistics
one then gets
\begin{equation}
(\Delta C_l)^2 \equiv \langle (C_l - \langle C_l \rangle_{ens})^2 \rangle
= \frac{2\langle C_l\rangle_{ens}^2}{2l+1}.
\label{e1}
\end{equation}
The variance $\Delta C_l$ thereby decreases with $1/\sqrt{2l+1}$.

The purpose of this paper is to present a first principle calculation
of the theoretical error in $C_l$ for the tensorial component of the CMB
anisotropy caused by a gravitational wave (GW) background.
We concentrate on the tensor contribution because after the Planck
era the GW couples extremely weakly to matter and is essentially a
free field propagating on a curved metric. During inflation, quantum
fluctuations in
the gravitational field are stretched out of the horizon where they
freeze and remain constant in amplitude. When they re-enter
the horizon much later they appear as classical GW's. These GW's are
often approximated as a classical stochastic background in the literature
$\cite{Wise2,White}$ and Eq.~($\ref{e1}$) is based on
this approximation. Using the Heisenberg representation approach
$\cite{Gris1}$, we do not make this
approximation and are able to calculate the cosmic variance explicitly.
As ours will be a quantum
mechanical calculation, we shall use $\langle {\cal O}\rangle$ to
denote the expectation value of an operator. In classical language,
this is equivalent to taking the ensemble average.

We assume a flat Robertson-Walker metric.
The quantized graviton field operator $h_{ij}$ may be written as
\begin{equation}
h_{ij}(\eta, \vec x) = \frac{A}{R}\sum_{\vec q}\sum^2_{s=1}
        \left[p^s_{ij}(\vec q)a^s_{\vec q}(\eta)e^{i\vec q\cdot\vec
        x}+\bar p^s_{ij}(\vec q){a^s_{\vec
        q}}^\dagger(\eta)e^{-i\vec q\cdot\vec x}\right]\>,
\label{e2}
\end{equation}
where $A$ is an overall constant whose value is not important for our
purposes. $\eta$ and $R$ are the conformal time and cosmic scale factor
respectively while $p^s_{ij}(\vec q)$ is the circular polarization
tensor for a GW in state $s$. $a_{\vec q}^s(\eta)$
and ${a_{\vec q}^s}^\dagger(\eta)$ are lowering and raising operators
in the Heisenberg representation with their evolution governed by the
hamiltonian $\cite{Gris1}$
\begin{equation}
H = \sum_{\vec q}\sum_{s=1}^2
        \left\{
                q{a^s_{\vec q}}^\dagger(\eta)a^s_{\vec q}(\eta) +
                q{a^s_{-\vec q}}^\dagger(\eta)a^s_{-\vec q}(\eta) +
               2\sigma(\eta)
                \left[
                {a^s_{\vec q}}^\dagger(\eta){a^s_{-\vec q}}^\dagger(\eta)-
                a^s_{\vec q}(\eta)a^s_{-\vec q}(\eta)
                \right]
        \right\}\>,
\label{e3}
\end{equation}
where $\sigma(\eta)=iR'/2R$ and
the prime denotes derivative with respect to $\eta$. Note that it is in
a simple quadratic Gaussian form. The Heisenberg
evolution equations, $i{a^s_{\vec q}}'= [a^s_{\vec q},H]$, $i{{a^s_{\vec
q}}^\dagger}'= -[{a^s_{\vec q}}^\dagger,H]$,
can be solved exactly using the Bogolubov transformation:
\begin{eqnarray}
a^s_{\vec q}(\eta)&=& u^s_q(\eta)a^s_{\vec q}(\eta_0) +
v^s_q(\eta){a^s_{-\vec q}}^\dagger(\eta_0)\>,
\nonumber \\
{a^s_{\vec q}}^\dagger(\eta) &=& \bar u^s_q(\eta) {a^s_{\vec
q}}^\dagger(\eta_0) + \bar v^s_q(\eta)a^s_{-\vec q}(\eta_0)\>,
\label{e5}
\end{eqnarray}
where $\eta_0$ is some initial time. Then ${u^s_q}'= -iqu^s_q +
\bar v^s_qR'/R$, ${v^s_q}' = -iqv^s_q + \bar u^s_qR'/R$,
with the initial conditions $u^s_q(\eta_0)=1$, $u^s_q(\eta_0)=0$ {}
$\cite{Gris2}$.

In the Schr\"oedinger approach found in $\cite{BirDav}$ the wave-function
itself is time dependent while the operators are time independent
with the field operator written as
\begin{equation}
h_{ij}(\eta, \vec x) = \frac{A}{R}\sum_{\vec q}\sum^2_{s=1}
        \left[p^s_{ij}(\vec q)a^s_{\vec q}(\eta_0)h^s_q(\eta)e^{i\vec
        q\cdot\vec x}+\bar p^s_{ij}(\vec q){a^s_{\vec
        q}}^\dagger(\eta_0)\bar h^s_q(\eta)e^{-i\vec q\cdot\vec
        x}\right]\>.
\label{e2a}
\end{equation}
To find the relationship between $h$ and the $u$-$v$ coefficients,
we use Eq.~($\ref{e5}$) in Eq.~($\ref{e2}$) and compare with
Eq.~($\ref{e2a}$), giving $h^s_q =u^s_q + \bar v^s_q$. Hence,
${h^s_q}''+(q^2-R''/R)h^s_q=0$ and the initial
conditions for $u^s_q$ and $v^s_q$ give $h^s_q(\eta_0)=1$ and
${h^s_q}'(\eta_0)=-iq$.

Following $\cite{Gris2}$, the classical temperature fluctuation
$\delta T/T$  of the CMB is now replaced by a field operator,
$\delta T/T =-\int_{e}^{r}d\Lambda h_{ij}'(\eta, \vec x)e^ie^j/2$,
where the lower (upper) limit of integration represents the point of
emission (reception) of the photon. Note that $a_{lm}$, $C_l$, and
$C(\theta)$ defined in Eq.~($\ref{ct}$) now become field
operators $\cite{Gris2}$ with
\begin{eqnarray}
\langle C(\theta) \rangle &=& \langle 0\vert \frac{\delta
T}{T}(\hat e_1)\frac{\delta T}{T}(\hat e_2) \vert 0\rangle=
\frac{1}{4}\int_{e}^{r}d\Lambda_1 \int_{e}^{r}d\Lambda_2\>
e_1^{i_1} e_1^{j_1} e_2^{i_2} e_2^{j_2}
\nonumber \\
   &{}& \frac{\partial\>\>}{\partial\eta_1}\frac{\partial\>\>}{\partial\eta_2}
        \langle 0\vert h_{{i_1}{j_1}}(\eta_1,\vec x_1)
   h_{{i_2}{j_2}}(\eta_2,\vec x_2)\vert 0\rangle\>,
\label{e9}
\end{eqnarray}
where $\vert 0\rangle$ is the vacuum state of the system at time $\eta_0$.
Using Eq.~($\ref{e5}$), we find
\begin{eqnarray}
G^{(2)}_{i_1j_1,i_2j_2}(x_1,x_2)&\equiv&
        \langle 0\vert h_{{i_1}{j_1}}(x_1)h_{{i_2}{j_2}}(x_2)\vert0\rangle
\nonumber \\
        &=&
        A^2
        \sum_{\vec q,s} p^s_{{i_1}{j_1}}(\vec q){\bar p}^s_{{i_2}{j_2}}
(\vec q) e^{i\vec q\cdot(\vec x_1-\vec x_2)}
        \left[\frac{u^s_q(\eta_1) + \bar v^s_q(\eta_1)}{R(\eta_1)}\right]
        \left[\frac{\bar u^s_q(\eta_2) + v^s_q(\eta_2)}{R(\eta_2)}\right] \>,
\nonumber \\
        &{}&
\label{e12}
\end{eqnarray}
and $\langle C(\theta)\rangle$ is dependent on the two point Green's
function for the field.

Notice that $G^{(2)}_{i_1j_1,i_2j_2}(x_1,x_2) = \overline
G^{(2)}_{i_2j_2,i_1j_1}(x_2,x_1)$. For a classical field, however,
one would expect $G^{(2)}_{i_1j_1,i_2j_2}(x_1,x_2) =
G^{(2)}_{i_2j_2,i_1j_1}(x_2,x_1)$. From
Eq.~($\ref{e12}$), this holds only when $u^s_q + \bar v^s_q$ is real.
Of course,
$h^s_q=u^s_q + \bar v^s_q$ will not, in general, be real given the
initial conditions for $u^s_q + \bar v^s_q$.
It can be shown, however, that if immediately after inflation there is a
transition to the radiation-dominated era, $h^s_q$
is essentially real for all relevent GW wavelengths
up to an overall $\vec q$-dependent constant phase factor after inflation
$\cite{Ng,Allen}$. This phase can be considered as a
random spatial phase which determines the location of the node of a
Fourier mode of the wave. It is randomly distributed since quantum
fluctuations during inflation assign equal probability to modes which
differ only by a spatial translation $\cite{Albrecht,Ng2}$.
Consequently, the usual ``scale-invariant'' initial condition for a GW
generated from inflation at the end of the inflationary era at
$\eta=0$, (up to an irrelevant overall constant phase) is given by
${h^s_q}/R=1$, $(h^s_q/R)'=0$ as $\eta\to0$. Subsequent temporal
evolution of the
wave can be well approximated by solving the equation of motion for
$h^s_q$ under these initial conditions $\cite{Ng1}$.
More importantly, the use of classical GW's after inflation in the
calculation of $\langle C_l\rangle$ is justified, and $\langle C_l\rangle$
is determined only by the temporal phase of the GW.

Using the identity from $\cite{White}$,
\begin{equation}
\int d\Omega \bar Y_{lm}(\Omega) e^{i\vec q\cdot \vec x} p^s_{ij}(\vec q)
e^ie^j = J_l(q \vert \vec x \vert) H^s_m\>,
\label{e13}
\end{equation}
where $H^s_m = (\delta^{+}_s - i \delta^{\times}_s)\delta^2_m +
(\delta^{+}_s + i \delta^{\times}_s)\delta^{-2}_m$, and
\begin{eqnarray}
J_l(q \vert \vec x \vert) = &\pi&
        \left(\frac{2l+1}{4\pi}
              \frac{(l+2)!}{(l-2)!}
        \right)^{1/2}
        \sum_n i^n(2n+1)j_n(q\vert\vec x\vert)
        \Bigg[
        \frac{\delta^n_{l-2}}{(2l-1)(2l+1)(l-3/2)}
\nonumber \\
        &{}&
        -\frac{2\delta^n_{l}}{(2l-1)(2l+3)(l+1/2)}
        +
        \frac{\delta^n_{l+2}}{(2l+1)(2l+3)(l+5/2)}
        \Bigg]\>,
\label{e15}
\end{eqnarray}
we obtain
\begin{equation}
\langle C_l\rangle = \sum_m \langle a^\dagger_{lm} a_{lm} \rangle =
\int_e^rd\Lambda_1\int_e^rd\Lambda_2
\frac{\partial}{\partial\eta_1}\frac{\partial}{\partial\eta_2}
g_{l}(\eta_1,\eta_2)\>,
\label{e16}
\end{equation}
where
\begin{equation}
g_{l}(\eta_1,\eta_2) =A^2\sum_{\vec q}
J_l(q\vert \vec x_1\vert)\bar J_{l}(q\vert \vec x_2\vert)
\frac{h^s_q(\eta_1)}{R(\eta_1)} \frac{h^s_q(\eta_2)}{R(\eta_2)}\>.
\label{e17}
\end{equation}
Note that in Eq.~($\ref{e17}$) we have taken $h^s_q$ as a real function
obeying the scale invariant initial condition.

Next, to calculate the variance in $C_l$ we note that $\langle C_l^2 \rangle$
depends on the four-point Green's function for the
field. Using Eq.~($\ref{e5}$),
\begin{eqnarray}
G^{(4)}_{i_1j_1,\dots,i_4j_4}(x_1,\dots,x_4)=
    &{}& G^{(2)}_{i_1j_1,i_2j_2}(x_1,x_2)G^{(2)}_{i_3j_3,i_4j_4}(x_3,x_4) +
\nonumber \\
          &{}&
         G^{(2)}_{i_1j_1,i_3j_3}(x_1,x_3)G^{(2)}_{i_2j_2,i_4j_4}(x_2,x_4) +
\nonumber \\
          &{}&
         G^{(2)}_{i_1j_1,i_4j_4}(x_1,x_4)G^{(2)}_{i_2j_2,i_3j_3}(x_2,x_3)\>.
\label{e18}
\end{eqnarray}
This form of factorization is expected from a Gaussian theory.
{}From this, it is then easy to show that
\begin{equation}
\langle (C_l - \langle C_l\rangle)^2 \rangle
=2\sum_{mm'}\vert\langle a^\dagger_{lm} a_{lm'}\rangle\vert^2.
\label{e19}
\end{equation}
{}From Eq.~($\ref{e13}$), it is easy to show that $\sum_s H^s_m \bar H^s_{m'}=
2\delta_{mm'}$, indicating that each component in $m$ is orthogonal to each
other. In addition, due to the isotropy of the GW background, equal
$\langle a^\dagger_{lm} a_{lm}\rangle$'s for a fixed $l$ would be expected.
Hence, it follows from Eqs.~($\ref{e16}$) and ($\ref{e19}$) that
\begin{equation}
(\Delta C_l)^2 = \frac{2\langle C_l\rangle^2}{2l+1},
\label{e20}
\end{equation}
which is exactly the classical result~($\ref{e1}$).

In conclusion, we have presented a calculation of the theoretical
error in $C_l$ based on quantum field theory.
We found from the isotropy and Gaussianity of the theory
that the fractional error in $C_l$ decreases with $1/\sqrt{2l+1}$,
as expected in the classical approximation.
We have, however, considered only the tensor contribution to the
anisotropy of the CMB in this paper.
As it is believed that the main contribution to the CMB
anisotropy is most probably due to the scalar perturbations, a
calculation of the cosmic variance for the scalar part is needed
before any definitive characterization of the experimental data can
be done with any certainty.

\begin{center}
{\bf Acknowledgements}
\end{center}

This work was supported in part by the R.O.C. NSC Grant Nos.
NSC84-2112-M-001-024 and NSC84-2112-M-001-022.

\end{document}